\documentclass[pra,twocolumn,superscriptaddress,floatfix,showpacs,10pt,letterpaper]{revtex4}
\usepackage{graphicx}
\usepackage{subfigure}
\usepackage{amssymb}
\usepackage{verbatim}
\usepackage{amsmath}
\usepackage{amscd}
\usepackage{amssymb}

\newtheorem{fact}{Fact}

\newtheorem{example}{Example}
\newtheorem{lemma}{Lemma}
\newtheorem{theorem}{Theorem}

\begin{document}

\title{Nonbinary Codeword Stabilized Quantum Codes}

\author{Xie Chen}
\affiliation{Department of Physics, Massachusetts Institute of
Technology, Cambridge, MA 02139, USA}

\author{Bei Zeng}
\affiliation{Department of Physics, Massachusetts Institute of
Technology, Cambridge, MA 02139, USA}

\author{Isaac L. Chuang}
\affiliation{Department of Physics, Massachusetts Institute of
Technology, Cambridge, MA 02139, USA}
\affiliation{Department of
Electrical Engineering, Massachusetts Institute of Technology,
Cambridge, MA 02139, USA}

\date{\today}


\begin{abstract}
The codeword stabilized (CWS) quantum codes formalism presents a
unifying approach to both additive and nonadditive quantum
error-correcting codes (arXiv:quant-ph/0708.1021), but only for binary states. Here we
generalize the CWS framework to the nonbinary case (of both prime and
nonprime dimension) and map the search for nonbinary quantum codes to a
corresponding search problem for classical nonbinary codes with 
specific error patterns. We show that while the additivity properties
of nonbinary CWS codes are similar to the binary case, the structural
properties of the nonbinary codes differ substantially from the binary case,
even for prime dimensions. In particular, we identify specific
structure patterns of stabilizer groups, based on which efficient
constructions might be possible for codes that encode more dimensions than any stabilizer codes of the same length and distance; similar methods cannot be applied in the binary
case. Understanding of these structural properties can help prune
the search space and facilitate the identification of good nonbinary CWS codes.
\end{abstract}

 \maketitle


\section{Introduction}
Classical error-correcting codes built on higher dimensional systems
may have better properties compared with binary codes and achieve
bounds that binary ones cannot reach \cite{MacWilliams}. For quantum
codes, similar improvements may also be expected when
error-correcting protocols built on qubit systems are extended to
the general qudit case \cite{Aharonov, Gottesman_qudit, Ashikhmin, Hamada, Rains} and many good qudit codes have indeed already been found \cite{Grassl}. The
recently developed codeword stabilized (CWS) quantum codes formalism
presents a unifying approach to construct both additive and
nonadditive binary quantum error-correcting codes \cite{CWS1, CWS2}.
Generalization of this framework to the nonbinary case could provide
a systematic means for constructing nonadditive, nonbinary quantum
codes. 

However, such a generalization is particularly challenging,
due to the change in group structure of the underlying error basis,
when moving to nonbinary quantum states. Such changes lead to
differences not only in the construction strategy, but also
structural properties of the codes. An open and interesting question
is whether these differences imply any possible advantage of
nonbinary CWS codes over their binary counterparts.

In this work, we generalize the CWS framework to the nonbinary
cases. Despite the fact that the group structure of error basis
changes dramatically with the dimension of the qudit, we show how
CWS codes can nevertheless be constructed for any dimension, prime or
nonprime, by mapping the search for nonbinary quantum codes 
to a corresponding search problem for nonbinary classical codes with
a specific error pattern. This mapping from quantum to classical for
constructing codes facilitates the search for quantum codes and
raises the hope that some good nonbinary classical codes may be used
to construct good nonbinary quantum codes, as has been done in the binary case by Grassl and Roetteler \cite{Markus1,Markus2}. This `classical' point of view also
helps understanding of the properties of CWS quantum codes. We show
that while the additivity properties of nonbinary CWS codes are
similar to the binary case, structural properties of the codes
differ substantially from the binary case, even for prime
dimensions.

These understandings provide essential clues in the search for codes with
better performance parameters. For example we show that the optimal
additive nonbinary quantum code, when mapped to a nonbinary
classical linear code, could be a subcode of a nonadditive quantum
code with the same distance, which is impossible in the binary case.
In the binary case, this structural property complicates any attempt
to search for codes beyond the stabilizer formalism. In the general
nonbinary case however, it is possible to find better codes by just
starting from the classical linear representation of the optimal
additive quantum code and adding codewords to it. Some other unifying frameworks similar to CWS have been proposed for
building both binary \cite{Yu2,Arvind,Calderbank} and higher
dimensional \cite{Yu3,Arvind,Looi} quantum codes. Based on these
frameworks, computer search has yielded promising results.
With further understanding of the structure properties of these code, we provide a basis for a systematic search for new classes of optimal nonadditive, nonbinary quantum codes.

The paper is organized as follows: Section II gives basic
definitions of generalized Pauli operations and qudit stabilizer
states, based on which, in Section III, a general construction for
nonbinary CWS code is given. Section IV is devoted to the discussion
of structure theorems of CWS codes, focusing especially on the
difference between the binary, prime, and nonprime dimensional
cases. Finally, we conclude in Section V.


\section{The {\em qudit} stabilizer states} In the qudit case, the
single qudit Pauli group $G_d^1$ for $d$ dimensional systems is generated
by $X,Z$ with the commutation
relation \cite{Bar1,San,Dab,Got,Pat,Kni}
\begin{equation}
ZX=q_dXZ, \label{qpc}
\end{equation}
where $q_d\equiv e^{i \frac {2\pi}{d}}$. Mathematically, it can be
proved that the group $G_d^1$ possesses a $d-$dimensional irreducible
representation. For general $d$, choose a basis
$\{|k\rangle\}_{k=0}^{d-1}$. We then have
\begin{equation}
Z|k\rangle=q_d^k|k\rangle, \, \ X|k\rangle =|k+1\rangle\, (k\in \mathbb{Z}_d).
\end{equation}

All the elements of the single qudit Pauli group are operators given by
$G_d^1=\{q_d^iZ^{j}X^{k},\, (i, j, k\in \mathbb{Z}_d)\}$. And the general commutation relation for any two elements is given by
\begin{equation}
q_d^{i_1}Z^{j_1}X^{k_1} \cdot q_d^{i_2}Z^{j_2}X^{k_2} = q_d^{j_1k_2-k_1j_2} \cdot q_d^{i_2}Z^{j_2}X^{k_2} \cdot q_d^{i_1}Z^{j_1}X^{k_1}
\end{equation}

We now define the $n$-qudit Pauli group. The familiar $n$-qubit Pauli group consists of all local operators of the form $R = \alpha_R R^{(1)}\dots R^{(n)}$, where $\alpha_R \in \{\pm 1, \pm i\}$ is an overall phase factor and $R^{(i)}$ is either the identity $I$ or one of the Pauli matrices $\sigma_x, \sigma_y$, or $\sigma_z$. We can define the analogous $n$-qudit Pauli group ${G}_d^n$ as the set of all local operators of the form $R = \alpha_R R^{(1)}\dots R^{(n)}$, where $\alpha_R = q_d^k$ for some $k \in \mathbb{Z}_d$ is an overall phase factor and $R^{(i)}$ is an element of the single qudit Pauli group $G_d^1$. $R$ can also be written in the form $R=\alpha_R' Z^{\mathbf{v}_R}X^{\mathbf{u}_R}$, where $\mathbf{v}_R$ and $\mathbf{u}_R$ are vectors over $\mathbb{Z}_d$ of length $n$, indicating the power of $Z$ and $X$ on each qudit. Define the weight of any Pauli operator $R$, denoted by $wt(R)$, as the number of qudits that $R$ acts nontrivially on. Then the set of operators in $G_d^n/\mathbb{Z}_d$ form a basis for all qudit
operators of weight $n$ or less. (For simplicity and without causing confusion, we will later on say that the $n$-qudit Pauli group $G_d^n$ forms a basis for all $n$-qudit operators.)

In $d$-dimensional systems, an $n$-qudit stabilizer state $|\psi\rangle$ is the
simultaneous eigenvector, with eigenvalues $1$, of a subgroup of $d^n$
commuting elements of $G_d^n$ which does not contain
multiples of the identity other than the identity itself. We call
this subgroup the \emph{stabilizer} $S$ of $|\psi\rangle$. A
minimal generating set for $S$ consists of $m$ elements. Note
that, as opposed to the situation for qubits, $m$ can be larger than
$n$. If $d$ has only single prime factors, then
$m=n$. If $d$ has multiple prime factors, then $n\le m\le 2n$ \cite{Moor3}. A
simple example for $d=4$ and $n=1$ is the state
$(|0\rangle+|2\rangle)/\sqrt{2}$ with stabilizer
$\{I,X^2,Z^2,X^2Z^2\}$: in this case $m=2$, $n=1$.


\section{General construction of qudit CWS code}
The qudit formalism presented in the previous section provides the
essential foundation for generalizing the CWS construction, to provide
a means for obtaining a quantum code $\mathcal{Q}$ for $d$-dimensional
states by mapping the code construction problem to identification of a
suitable a classical code $\mathcal{C}$.
An important observation in quantum error correction is that a code can correct a linear space of continuous errors  if it can correct the finite set of `basis' error operators of the space \cite{Nielsen}. The $n$-qudit Pauli group $G_d^n$ forms a basis for all $n$-qudit operators and will serve as the `error basis' in our discussion. Special attention should be paid to
this error basis when moving from binary to nonbinary, especially
nonprime dimensional systems, as its group structure changes
dramatically. Although this could lead to changes in the form of
representation of CWS codes, we nevertheless show how CWS codes can
be constructed in the most general case. We denote by $\mathcal{Q}=((n,K,\delta))_d$ a CWS code built
on $n$ qudits which encodes a $K$ dimensional logical space and can detect any error of weight less than $\delta$. In particular, an additive code encoding $k$ logical qudits in $n$ physical qudits which can detect any error of weight less than $\delta$ is denoted as $[[n,k,\delta]]_d$.

An $((n,K,\delta))_d$ CWS code of qudits is
described by two objects: $S$, an $n$-qudit stabilizer subgroup of
the qudit Pauli group, which we call the `word
stabilizer', together with a family of $K$ $n$-qudit Pauli operators,
$W=\{w_l\}_{l=1}^K$, which we call the `word operators', following the terminology in the binary case \cite{CWS1}.

$S$ acts on $n$ qudits and can be generated by $m$ Pauli operators,
$S=\langle g_1,g_2,...,g_m\rangle$. Note that while $m = n$ for qubit systems, for qudits, in general, $m \ge n$. This is a major change in the group structure of $S$ when moving from binary to nonbinary case.  It is noted in \cite{Moor3} that if $m>n$, the imposed condition in Ref. \cite{S2} for a stabilizer state to be equivalent to a graph state, is not fulfilled. Therefore we will no longer transform the stabilizer state into a graph state to obtain a standard form of CWS code as it was done in the binary case \cite{CWS1}. Instead, we will just base
the construction on general stabilizer states, to connect to the
classical error correction theory used in the nonbinary CWS framework.

Similar to the binary case, $S$ can be represented in the
format $[r|t]$, where $r,t$ are $m\times n$ matrices and the $i$th
row $j$th column element is the exponent of $X$(or $Z$) operator on
the $j$th qudit in the $m$th generator. There is a unique state
$|S\rangle$ stabilized by $S$, i.e. $|S\rangle$ satisfies
$s|S\rangle=|S\rangle$ for all $s\in S$. The $((n,K,\delta))_d$ CWS code $\mathcal{Q}$ is the space spanned by
basis vectors of the form $|w_l\rangle=w_l|S\rangle$, where $w_l \in W, l=1...K$
are word operators.


\subsection{Mapping of quantum basis vectors to classical strings}
Note each $|w_l\rangle$ is stabilized by $w_lSw_l^{\dagger}$ and for $g_k \in  S$,
$w_lg_kw_l^{\dagger}=q_d^{l_k}g_k$, where $q_d=e^{i2\pi/d}$. So each $|w_l\rangle$ can be represented by classical string over $\mathbb{Z}_d$, of the form
\begin{equation}
\mathbf{c}_l=(l_1,l_2,...,l_m),
\end{equation}
i.e. each $|w_l\rangle$ is stabilized by $w_lSw_l^{\dagger}=\langle
q_d^{l_1}g_1,q_d^{l_2}g_2,...,q_d^{l_m}g_m\rangle$. Without loss of
generality, we choose the first basis state to be stabilized
by $S$, i.e. $\mathbf{c}_0 = (0,0,\dots,0)$. The set of basis
vectors $\{|w_l\rangle\}$ can now be specified completely, up to a phase, by word stabilizer $S$ and
classical codewords $\mathcal{C} = \{\mathbf{c}_l\}$. This step is analogous to the construction in binary case \cite{CWS1}.

\textbf{Remark 1}: This mapping from stabilizer states to classical strings is not unique. When the generating set for $S$ is chosen differently, the same stabilizer state $|w_l\rangle$ could correspond to different classical strings. Suppose generating set $\langle g_1,g_2,...,g_m\rangle$ is transformed into $\langle g'_1,g'_2,...,g'_{m'}\rangle$ via an $m\times m'$ matrix $R$ over $\mathbb{Z}_d$ as $g'_t=\prod\limits_{k=1}^{m}g_k^{R_{kt}}$. Note that $w_lg'_tw^{\dagger}_l = w_l\prod\limits_{k=1}^{n}g_k^{R_{kt}}w^{\dagger}_l = (q_d^{\sum_{k=1}^{n}l_kR_{kt}})g'_t = q_d^{l'_t}g'_t$. Therefore the corresponding classical strings transform as $l'_t = \sum_{k=1}^{n}l_kR_{kt}$. If $\mathcal{C}$ is a matrix having the classical codewords $\mathbf{c}_l$'s as its row vectors,
\begin{equation}
\mathcal{C}' = \mathcal{C}R
\end{equation}

\textbf{Remark 2}: The multiplication of codeword operator $w_l$'s is equivalent to the addition of corresponding classical string $\mathbf{c}_l$'s as can be seen from
\begin{equation}
w_mw_lg_kw_l^{\dagger}w_m^{\dagger}=q_d^{l_k}w_mg_kw_m^{\dagger}=q_d^{l_k+m_k}g_k
\end{equation}


\subsection{Mapping of quantum errors to classical strings}

Based on the above mapping from stabilizer states to classical strings, we can
show that in this CWS formalism, quantum errors also have a classical representation, just as in the binary case
\cite{CWS1}. As discussed before, we will only focus on $n$ qudit Pauli error operators of the form $E = \alpha_E Z^{\mathbf v}X^{\mathbf u} = \alpha_E Z_1^{v_1}...Z_n^{v_n}X_1^{u_1}...X_n^{u_n}$ where $\alpha_E = q_d^k, k\in\mathbb{Z}_d$ and $\mathbf{v},\ \mathbf{u}$ are vectors over $\mathbb{Z}_d$. This is because $n$ qudit Pauli operators form a basis of all $n$ qudit errors. The action of $E$ on a code basis state $|w_l\rangle$ takes it to $|w_l^E\rangle = E|w_l\rangle$, which is still a stabilizer state, and the corresponding classical codeword changes from $\mathbf{c}_l$ to $\mathbf{c}_l'$ which is equivalent to the addition (mod $d$) of classical string $Cl_S(E)$ to $\mathbf{c}_l$. Hence $E$ has a classical representation $Cl_S(E)$ and in particular, we prove that:

\begin{theorem}
\begin{equation}
Cl_S(E=\alpha_E Z^{\mathbf v}X^{\mathbf u})=\sum\limits_{l=1}^n v_l\mathbf{r}_l+u_l\mathbf{t}_l,
\end{equation}
where $\mathbf{r}_l$ and $\mathbf{t}_l$ are the $l$th column of the matrices $r$ and $t$ respectively, which are all length $m$ strings over $\mathbb{Z}_d$.
\end{theorem}
\textbf{Proof}: The action of error $E$ on a codeword state $|w_l\rangle$ is
$|w_l^E\rangle = Ew_l|S\rangle$. This state is stabilized by
$Ew_lg_kw_l^{\dagger}E^{\dagger} = q_d^{l_k}Eg_kE^{\dagger}$

Let
\begin{equation}
E = \alpha_E Z_1^{v_1}Z_2^{v_2}...Z_n^{v_n}X_1^{u_1}X_2^{u_2}...X_n^{u_n}
\end{equation}

Let $r$ and $t$ be matrices representing $X$ and $Z$ part of $S$ respectively. $g_k$ contains $X_1^{r_{k1}}$. Conjugating $g_k$ by $E$ will
result in a prefactor of $q_d^{r_{k1}\cdot v_1}$. The same argument
applies to $Z$ factors and to other coordinates. All these
prefactors add up and the stabilizer of state $|w_l^E\rangle$ is
given by
\begin{equation}
Ew_lg_kw_l^{\dagger}E^{\dagger} = q_d^{l_k}q_d^{v_1\cdot r_{k1} + ...+ v_n \cdot r_{kn} + u_1 \cdot t_{k1} + ...+ u_n \cdot t_{kn}}g_k
\end{equation}

Then the $k$th element in the corresponding classical codeword is changed by
\begin{equation}
v_1\cdot r_{k1} + ...+ v_n \cdot r_{kn} + u_1 \cdot t_{k1} + ...+ u_n \cdot t_{kn}
\end{equation}

Therefore, the change to each codeword induced by $E$ is
\begin{equation}
Cl_S(E)= \sum\limits_{l=1}^n v_l\mathbf{r}_l+u_l\mathbf{t}_l
\end{equation}$\Box$

\subsection{Error correction condition}

Recall that an $((n,K))_d$ quantum error-correcting code is a $K$ dimensional subspace (over the complex field) of the $n$-qudit Hilbert space $\mathcal{H}_d^{\otimes n}$. Given an orthonormal basis $\{{|\psi_i\rangle}\}_{i=0}^{K-1}$ of such a subspace, the code detects errors in the set $\{E_a\}_{a=1}^r$, if and only if the error correction condition
\begin{equation}
\langle{\psi_i}| E_a |{\psi_j}\rangle = C_a \delta_{ij}
\label{eq-condition}
\end{equation}
is satisfied for all $i, j \in \{0...K-1\}$ and all $a \in \{1...r\}$ and
$C_a$ is independent of $i$ and
$j$~\cite{Nielsen,knill-laflamme-theory,bennett-tome}. A code for which $C_a = 0, \forall a$ is called a {\em nondegenerate} code, while a code for which this is not true is called {\em degenerate}.  

With the above mapping, one can
see that the basis vectors $|w_l\rangle$'s will not be
taken into each other by error $E$ as long as the corresponding
classical strings $\mathbf{c}_l$'s are not mixed by addition (mod $d$) of $Cl_S(E)$.
However, to ensure that the code corrects quantum errors, one also has to check conditions that are quantum--that superpositions of basis vectors are not mixed either \cite{Nielsen}. 
Therefore for CWS code $\mathcal{Q}$ spanned by $\{|w_l\rangle\}_{l=1}^K$ to be able to detect
errors from set $\mathcal{E} = \{E\}$, $\{\mathbf{c}_l\}$ should be
able to detect errors from $\{Cl_S(E)\}$ and:
\begin{eqnarray}
\forall E\ \in \mathcal{E}, Cl_S(E)\ne 0 \nonumber\\
\text{\ or\ } \forall l, w_lE = Ew_l
\label{ECC}
\end{eqnarray}

If $\exists E$ such that $Cl_S(E) = 0$, $E$ must act trivially on the codespace, i.e. the code is degenerate. By mapping the basis vectors and the error patterns all to classical strings, we hope to reduce the construction of quantum codes to that of classical ones. However, it seems that the existence of degenerate quantum codes prevents such a simple reduction and both the classical and quantum (Equation \ref{ECC}) part of the error correcting condition have to be taken into consideration.

However, this non-classical problem can be fixed by choosing properly the generators of a given stabilizer $S$. We show that there exists a canonical form of the corresponding classical codewords which automatically takes care of the degeneracy of the code and hence only classical error correction conditions need to be checked when searching for new codewords using this form of representation.

First identify all errors $E$ that belong to $S$ (up to some coefficient). In particular, we consider an $((n,K,\delta))$ CWS code which detects all errors with weight less than $\delta$. Take all weight less than $\delta$ elements from ${S}$ and form a set ${S}_{\delta}$(which may not form a group). If the rank of ${S}_{\delta}$ is $r$, then we can choose $r$ independent elements $g_1, ...,g_r \in {S}_{\delta}$ and $g_{r+1},...,g_m \notin {S}_{\delta}$ such that $\langle g_1,...,g_r,g_{r+1},...,g_m\rangle$ generate ${S}$. According to the above discussion, on this basis, the classical codewords must all be $0$ on the first $r$ coordinates. We call this `the canonical form' of degenerate codes.

If we are only searching for new codewords by linear combination of existing classical codewords, it is easy to see that the new codeword states are still eigenvalue $1$ eigenstates of all $g_i \in {S}$ with $wt(g_i) < \delta$ and the degeneracy condition is automatically satisfied. This spares us the work of checking the quantum part of the error correction condition and allows us to focus our attention on the classical part. In the following discussion, we will always assume that this `canonical form' has been taken.

Now we have all the necessary elements (summarized in Table \ref{QvC}) to construct a $((n,K,\delta))_d$ nonbinary CWS quantum codes from classical $(m,K)$ code over $\mathbb{Z}_d$. This proceeds in a similar way to the binary case \cite{CWS2}. First choose a `word stabilizer' $S$ and determine $Cl_S(E)$ for all $E \in \mathcal{E}$. (Which `word stabilizer' could lead to better codes is unknown, even in the binary case. As we show below, we give some hint on specific structural properties of $S$, based on which search for good CWS codes might proceed more efficiently.) Without loss of generality, the first classical codeword $\mathbf{c}_0$ is chosen to be $(0,0,...,0)$. Next, check the `quantum part' of the error correction condition (i.e. Equation \ref{ECC}) and exclude inadmissible classical codewords. The rest length $m$ classical strings in $\mathbb{Z}_d$ form a `CWS clique graph' whose vertices are classical codewords and whose edges indicate codewords that can be in the same classical code together. Finding the maximum clique in graphs based on all possible choice of $S$ gives the CWS code with optimal $K$ for fixed $n$, $\delta$, and $d$, similar to the binary case \cite{CWS2}.

\begin{table}
\centering
\begin{tabular}{|c|c|c|}
\hline
$((n,K,\delta))_d$ & Quantum & Classical \\
CWS code & Representation & Representation\\
\hline
Basis & State $|w_l\rangle$ stabilized by & Length $m$ string over $\mathbb{Z}_d$\\
State & $\langle q_d^{l_1}g_1,...,q_d^{l_m}g_m\rangle$ & $\mathbf{c}_l = (l_1,...,l_m)$\\
\hline
Error & Qudit Pauli Operator & Length $m$ string over $\mathbb{Z}_d$ \\
Pattern & $E=\alpha_E Z^{\mathbf{v}}X^{\mathbf{u}}$& $Cl_S(E)=\sum\limits_{l=1}^n v_l\mathbf{r}_l+u_l\mathbf{t}_l$\\
\hline
Error & $\langle w_i|E|w_j\rangle = c_E \delta_{ij}$ & $\{\mathbf{c}_l\}$ detects $\{Cl_S(E)\}$\\
Detection & $\forall i,j$ & and $\forall E, Cl_S(E)\ne 0$ \\
Condition & and $\forall E$ & or $\forall l, w_lE = Ew_l$ \\
\hline
\end{tabular}
\caption{Summary: Quantum and Classical representation of an $((n,K,\delta))_d$ CWS code. The classical representation is based on a particular choice of generating set $\left(g_1,...,g_m\right)$ for the $n$-qudit stabilizer group $S$. $\alpha_E = q_d^k, k\in\mathbb{Z}_d$ and $\mathbf{v},\ \mathbf{u}$ are length $n$ vectors over $\mathbb{Z}_d$. $Cl_S(E)$ is the classical representation of error $E$ based on $S = \langle g_1,...,g_m\rangle$. $\mathbf{r}$ and $\mathbf{t}$ are matrices representing the $X$ and $Z$ part of $S$ respectively and $\mathbf{r}_l$ and $\mathbf{t}_l$ are their $l$th column.}
\label{QvC}
\end{table}


\section{Structure theorems}
The above mapping from quantum to classical as given in Table \ref{QvC} enables us to search for CWS codes on $n$-
qudit quantum systems by searching through the
corresponding classical codeword space of $m$ $d$-dimensional
classical systems. This is still a hard task. Suppose we wish to
find the maximum $K$ given $n$, $\delta$ and $d$. As explained above, we can encode this problem into that of finding a maximum clique in a `CWS clique graph'. Given $S$ and $\delta$, the `CWS clique graph' has $~d^m$
vertices and searching for the maximum clique of it takes
exponential time. There are also exponentially many word stabilizers $S$ to consider.

A highly desired insight, which could help prune the search space,
would be knowledge about how the structure of the stabilizer subgroup $S$ and the
classical codewords $\mathcal{C}$ might affect the properties (size, additivity, etc.) of
the CWS code constructed. In the binary case, several
structure theorems about this relationship were given \cite{CWS2}. In
particular, the structure theorems given for the binary case provided insight into how and whether prior knowledge about good
codes might help reduce the complexity in the search for better
codes. For example, one may start from the classical representation
of the optimal additive code for given $n$, $\delta$, $d$ and expand the code space by adding more classical
codewords, in hope of obtaining a larger coding space. However, as
a structure theorem (Corollary 4 in \cite{CWS2}) shows, if the corresponding classical
representation is linear, the code space cannot be enlarged.

In this section, we expand the discussion about structural properties of CWS codes to the general qudit
case, not only for prime $d$ but any composite number as well. For
each structure theorem given for the binary case \cite{CWS2}, we either prove similar theorems for the nonbinary case or give explicit counter-example to it. As the properties of classical codes change when $d$ moves from binary to nonbinary, prime to nonprime, the properties of the corresponding CWS codes change as well. Interestingly,
we find that the restriction on expanding linear classical codeword
spaces can be lifted in some cases. This implies a possible short
cut for finding CWS codes which encode larger logical spaces than any stabilizer code with same $n$ and $\delta$.

We first list all the structure theorems in the binary case
and then examine the corresponding situations of general $d$. $d$
being prime or not can have big difference on the structure
properties of CWS codes and we will separate these two cases in our
discussion.

Before proceeding, we first note that the additivity property of binary CWS codes can naturally be generalized to arbitrary $d$.
\begin{fact}
If the classical codewords $\mathcal{C}$ of CWS code $\mathcal{Q}$ are linear, then $\mathcal{Q}$ is an additive quantum code.
\end{fact}

\begin{fact}
If CWS code $\mathcal{Q}$ is additive, then there exists an $S$ and a linear $\mathcal{C}$ which defines $\mathcal{Q}$.
\end{fact}

Note that the classical codewords $\mathcal{C}$ defining an additive
CWS code $\mathcal{Q}$ could be nonlinear, for some choice of $S$.
The proof of the above facts proceeds in a similar way as has been done
for the binary case \cite{CWS1} and can be found elsewhere
\cite{Arvind}. A similar proof based on qudit graph states can be found in \cite{Looi}.

\subsection{The binary case}When $d = 2$, the following theorems hold:
\begin{theorem}
All $((n,2,\delta))_2$ CWS codes are additive.
\end{theorem}

\begin{theorem}
Any $((n,3,\delta))_2$ CWS codes is a subcode of some $[[n,2,\delta]]_2$ additive code.
\end{theorem}

We say an additive code is optimal for given $n$ and $\delta$ if it can encode $2^k$ dimensions and there does not exist another additive code which can encode $2^{k+1}$ dimensions.

\begin{theorem}
The linear classical codewords $\mathcal{C}$ representing the optimal additive code $((n, 2^k, \delta))_2$ cannot be a subcode of the classical codewords $\mathcal{C}'$ of another CWS code with parameters $((n, K > 2^k, \delta))_2$.
\end{theorem}

Hence we see that in the binary case, we cannot start from a linear classical representation of optimal additive code and obtain larger coding space by adding classical codewords.

\subsection{The prime dimension case}When $d$ is prime, we can prove a theorem similar to the binary case
\begin{theorem}
Any $((n,2,\delta))_d$ CWS code is a subcode of some $[[n,1,\delta]]_d$ additive code, when $d$ is prime.\label{n2d}
\end{theorem}
\textbf{Proof}: WLOG we can choose the first codeword to be $\mathbf{c}_0 = (0,0,\dots,0)$. The second codeword is $\mathbf{c}_1 = (i_1, \dots, i_n)$, $i_k \in \mathbb{Z}_d$. Any error with weight less than $(\delta-1)/2$ is of the form
\begin{equation}
Cl_S(E=\alpha_E Z^{\mathbf{v}}X^{\mathbf{u}})=\sum\limits_{l=1}^n v_l\mathbf{r}_l+u_l\mathbf{t}_l,
\end{equation}
where $wt(\mathbf{v} + \mathbf{u}) \leq (\delta-1)/2$ (Note addition here is not mod $d$, as $\mathbf{v}$ and $\mathbf{u}$ correspond to $Z$ and $X$ error respectively). $r$ and $t$ are matrices representing the $X$ and $Z$ part of $S$ respectively.

Now we want to prove that $2\mathbf{c}_1$ is also a codeword. If this is not true, then there exists some error with weight $\leq (\delta-1)/2$, s.t.
\begin{equation}
2\mathbf{c}_1 = \sum\limits_{l=1}^n v_l\mathbf{r}_l+u_l\mathbf{t}_l
\end{equation}

As $d$ is a prime, $2$ and $d$ are coprime. There exists some integer $q$ s.t. $2q = 1$ mod $d$. Therefore
\begin{equation}
2q\mathbf{c}_1 =\mathbf{c}_1 = \sum\limits_{l=1}^n qv_l\mathbf{r}_l+qu_l\mathbf{t}_l
\end{equation}

Multiplication by $q$ will not increase the weight of the error which means that $\mathbf{c}_1$ cannot be a codeword of distance $\delta$. This is contradictory to our assumption, therefore $2\mathbf{c}_1$ must also be a codeword.

In the same way we can show that $3\mathbf{c}_1,\dots, (d-1)\mathbf{c}_1$ can all be added to the original codeword set and they form a group. Then we have an $[[n,1,\delta]]_d$ additive code.$\square$

In correspondence to the second structure theorem in the binary case, we find that when $d > 2$ it is not always possible to add a fourth dimension to a 3-dimensional $((n,3,\delta))_d$ code in a similar way to the binary case.
\begin{example}
Consider the 7-qutrit ($d = 3$) stabilizer state with $S = \langle Z_{i-1(mod\ 7)}X_iZ_{i+1(mod\ 7)}, i = 0...6\rangle$. The first three codewords for a distance 3 code can be chosen as $\vec{0}$, $\mathbf{c}_1 = (1 1 0 0 1 0 0)$, $\mathbf{c}_2 = (0 0 1 0 0 1 1)$. Distance between any two of these three codewords is at least 3. But if we want to add a fourth one, for example $\mathbf{c}_1 - \mathbf{c}_2 = (1 1 2 0 1 2 2)$ to the codeword set, it is only of distance $2$ from $\mathbf{c}_2$, as $\mathbf{c}_1-2\mathbf{c}_2 = (1 1 1 0 1 1 1) = Cl_S(Z_1Z_5X_1X_5)$. Similarly $\mathbf{c}_1+\mathbf{c}_2$ is not a valid codeword either, as $\mathbf{c}_1+\mathbf{c}_2 = (1 1 1 0 1 1 1) = Cl_S(Z_1Z_5X_1X_5)$.
\end{example}
Therefore in general we can not always expand a 3-dimensional code
into a 4-dimensional one by linear combination of existing classical
codewords.

\begin{theorem}
The linear classical codewords $\mathcal{C}$ over $\mathbb{Z}_d$ representing the optimal additive code $((n, d^k, \delta))_d$ cannot be a subcode of the classical codewords $\mathcal{C}'$ of another CWS code with parameters $((n, K > d^k, \delta))_d$, when $d$ is prime. \label{a2n}
\end{theorem}
\textbf{Proof}: Suppose the classical codewords for the optimal additive code are $(\mathbf{0}, \mathbf{c}_1, \mathbf{c}_2, \dots, \mathbf{c}_{d^k-1})$. If we can add one more dimension $\mathbf{c}_{d^k}$ to the set without changing the distance of the code, we can actually show that the whole set
\begin{eqnarray}
(\mathbf{0}, \mathbf{c}_1, \dots,  \mathbf{c}_{d^k-1},
\mathbf{c}_{d^k}, \mathbf{c}_{d^k}-\mathbf{c}_1, \dots, \mathbf{c}_{d^k}-\mathbf{c}_{d^k-1}, \nonumber \\
2\mathbf{c}_{d^k}, 2\mathbf{c}_{d^k}-\mathbf{c}_1, \dots, 2\mathbf{c}_{d^k}-\mathbf{c}_{d^k-1},
\dots, \nonumber \\
(d-1)\mathbf{c}_{d^k}, (d-1)\mathbf{c}_{d^k}-\mathbf{c}_1, \dots, (d-1)\mathbf{c}_{d^k}-\mathbf{c}_{d^k-1})
\end{eqnarray}
is an $((n, d^{k+1}, \delta))$ additive code. The codewords form a group so it only remains to check that they satisfy the error correction condition.

The difference between any two codewords is of the form $q\mathbf{c}_{d^k}-\mathbf{c}_j$, $q = 0,\dots,(d-1)$, $j = 0, \dots, (d^k-1)$. $q$ and $d$ are coprime, so $\exists s$ s.t. $sq = 1 \  \text{mod}\  d$.  Suppose $q\mathbf{c}_{d^k}-\mathbf{c}_j$ is of weight $f$ and $q\mathbf{c}_{d^k}-\mathbf{c}_j = \sum\limits_{l=1}^n v_l\mathbf{r}_l+u_l\mathbf{u}_l$, for some $v_l$ and $u_l$. Multiply both sides by $s$ we have $\mathbf{c}_{d^k}-s\mathbf{c}_j = \mathbf{c}_{d^k}-\mathbf{c}_{j'} = \sum\limits_{l=1}^n sv_l\mathbf{r}_l+su_l\mathbf{t}_l$, so the distance between $\mathbf{c}_{d^k}$ and $\mathbf{c}_{j'}$ is bounded above by $f$. However we know the distance is bounded below by $\delta$, therefore $\delta \leq f$ and any two codewords in the above set satisfy the error correction condition for distance $\delta$.$\square$

\subsection{The nonprime dimension case} If $d$ is nonprime, then not
all $q < d$ have an inverse on the ring over $\mathbb{Z}_d$
and some of the above arguments no longer hold.

For example, the attempt to expand any $((n,2,\delta))_d$ CWS code into an $[[n,1,\delta]]_d$ additive code by linear combination of classical codewords may fail as shown by the following example.
\begin{example}
Let $d = 4$. Consider the stabilizer group $S$ represented in the matrix form $[r|t]$ where

$ r = \begin{pmatrix}1 & 0 & 0 \\0 & 1 & 0\\ 0 & 0 & 1 \end{pmatrix}$,
$ t = \begin{pmatrix}0 & 2 & 2 \\ 2 & 0 & 0 \\2 & 0 & 0 \end{pmatrix}$.

We can check that $\mathbf{c}_1 = (011)$ is of distance $2$ from
$\mathbf{0}$ but $2\mathbf{c}_1 = (022) = Cl_S(Z_1)$.
\end{example}
In general $q \mathbf{c}_1$ might have less weight than $\mathbf{c}_1$ and
the extension used in Theorem \ref{n2d} breaks down for nonprime
dimensions.

On the other hand, the reasoning for Theorem \ref{a2n} also breaks
down. This actually opens up the possibility of finding a good CWS
code starting from the linear representation of the best known
additive code. For this to be possible, the stabilizer group $S$
should have certain structure. Define the Greatest Common Divisor
(GCD) between a column vector $\mathbf{v} = (v_1, \dots, v_n)$ and a
number $p$ to be the greatest common divisor of the set $(v_1,
\dots, v_n, p)$. In our problem, the column vector could be a
codeword or an error pattern. If the GCD between $\mathbf{v}$ and $p$
is 1, we say that $\mathbf{v}$ and $p$ are coprime. If the GCD is
strictly greater than 1, we say that they are not coprime. We prove
the following theorem:

\begin{theorem}
The linear classical code representing an optimal additive $((n,
d^k, \delta))_d$ code can be a subcode of the classical
representation of another $((n, K>d^k, \delta))_d$ CWS code only if
the underling `word stabilizer' $S$ has a weight less than $\delta$ error pattern
$Cl_S(E) = \sum\limits_{l=1}^n v_l\mathbf{r}_l+u_l\mathbf{t}_l$ which
satisfies $GCD(Cl_S(E), d) = m >1$ and $m$ does not divide $GCD(v_1, \dots, v_n,
u_1,\dots, u_n, d)$.
\end{theorem}

First we prove the following lemma:
\begin{lemma}
Suppose we can add one more dimension $\mathbf{c}_{d^k}$ to a linear
classical code $\mathcal{C}$ representing additive code $((n, d^k,
\delta))_d$ and form a $((n, d^k+1, \delta))_d$ code. For any $q \in
\mathbb{Z}_d$, codewords $q\mathbf{c}_{d^k}-\mathbf{c}_j, j =
0,\dots,d^k-1$ must have weight larger than $\delta$ unless $GCD(q,
d)= m > 1$ and $GCD(m, \mathbf{c}_j) = m$.
\end{lemma}
\textbf{Proof}:1. If $GCD(q,d) = 1$, we can prove in a way similar
to the prime dimension case that there exists certain $j' \in
\{0,\dots,d^k-1\}$ such that $\mathbf{c}_{d^k}-\mathbf{c}_{j'}$ has weight
no more than that of $q\mathbf{c}_{d^k}-\mathbf{c}_j$. However
$wt(\mathbf{c}_{d^k}-\mathbf{c}_{j'})$ is bounded below by $\delta$,
therefore $wt(q\mathbf{c}_{d^k}-\mathbf{c}_j) \geq \delta$, where $wt(\mathbf{v})$ is the Hamming weight of vector $\mathbf{v}$.

2. If $GCD(q,d)= m > 1$ but $GCD(m, \mathbf{c}_j) < m$, suppose $s$ is
the smallest integer that satisfies $sm = 0$ mod $d$. Multiplying
$q\mathbf{c}_{d^k}-\mathbf{c}_j$ by $s$ we get some codeword $\mathbf{c}_l$
from the original additive code which is not $0$. The fact that
$\mathbf{c}_l$ has weight larger than $\delta$ ensures that
$wt(q\mathbf{c}_{d^k}-\mathbf{c}_j) > \delta$.

So for the codeword $q\mathbf{c}_{d^k}-\mathbf{c}_j$ to have weight less
than $\delta$, both conditions must be satisfied, i.e. $GCD(q, d)= m
> 1$ and $GCD(m, \mathbf{c}_j) = m$. $\Box$

\textbf{Proof of theorem}:

It is possible for a linear $\mathcal{C} = (\mathbf{0}, \dots,
\mathbf{c}_{d^k-1})$ of an additive code  to be a subcode of
$\mathcal{C}' = (\mathbf{0}, \dots, \mathbf{c}_{d^k-1}, \mathbf{c}_{d^k})$ of
another CWS code  but not a larger additive code only when for some
$q \in \mathbb{Z}_d$ and $j \in \{\mathbf{0},\dots,d^k-1\}$ some
codewords $q\mathbf{c}_{d^k}-\mathbf{c}_j$ has weight less than $\delta$.

We know from the lemma, this implies that
$GCD(q\mathbf{c}_{d^k}-\mathbf{c}_j, d) = m
> 1$ which means that there exists an error pattern $Cl_S(E) = q\mathbf{c}_{d^k}-\mathbf{c}_j$
of weight less than $\delta$ and $GCD(Cl_S(E), d) = m >1$.

Suppose $Cl_S(E) = \sum\limits_{l=1}^n v_l\mathbf{r}_l+u_l\mathbf{t}_l =
q\mathbf{c}_{d^k}-\mathbf{c}_j$. If $m$ can divide both $v_l$ and $u_l$, $\forall l \in \{
1,\dots, n\}$, divide both sides by $m$ we get $\sum\limits_{l=1}^n
v_l'\mathbf{r}_l+u_l'\mathbf{t}_l = q'\mathbf{c}_{d^k}-\mathbf{c}_{j'}$.
$q'\mathbf{c}_{d^k}-\mathbf{c}_{j'}$ is then another codeword of weight
less than $\delta$. However $q'$ must be coprime to $d$, which contradicts the Lemma. Therefore, $m$ does not divide $GCD(v_l, u_l,
d)$.$\square$

This theorem specifies a special kind of stabilizer group $S$, based
on which codes exceeding all known additive ones might be found more
efficiently. This is a special property of nonprime dimension
systems and could lead to the discovery of better coding parameters.


\section{Conclusion and Discussion} Using the qudit Pauli group as a basis for describing quantum states and quantum errors, we have generalized the CWS framework to nonbinary quantum
systems. By mapping the code space basis vectors and the error
patterns to classical strings, we showed how the problem of
constructing nonbinary quantum error correcting codes and studying their
properties can be reduced to a corresponding problem for classical error correction codes. This reduction provides a systematic way for constructing good nonbinary quantum codes from known good nonbinary classical codes. We discussed also, using this classical point of view,
the properties of nonbinary CWS codes and found that while their
additivity properties are similar to the binary case, their
structure differs substantially from the binary case, even for prime
dimension $d$. The breakdown of some structure theorems in the qudit
case opens up possibilities for more efficient construction of codes
encoding more dimensions than the best known additive codes  and we identified structural properties
of the `word stabilizers' $S$ which enable this possibility. This
might give a short-cut for finding good nonbinary nonadditive codes. An important question which remains un-answered is the relation between the structure of $S$ and the maximum size of CWS code built on it. A complete understanding of this connection could save exponential time in the search for CWS codes and achieve major progress in the study of quantum error correction and fault-tolerant quantum computation.

\section*{Acknowledgment}

We thank Andrew Cross, Markus Grassl, Graeme Smith, and John Smolin for helpful discussions.


\begin{thebibliography}{10}

\bibitem{MacWilliams} F. J. MacWilliams and N. J. A. Sloane, \textit{The Theory of Error-Correcting Codes}. Amsterdam: North-Holland Publishing Company, (1977).

\bibitem{Aharonov} D. Aharonov and M. Ben-Or, Proceedings of the 29th Annual ACM Symposium on Theory of
Computing, Association for Computing Machinery, 176, (1997).

\bibitem{Gottesman_qudit} D. Gottesman, Lect. Notes. Comp. Sci., \textbf{1509}, 302 (1999).

\bibitem{Ashikhmin} A. Ashikhmin and E. Knill, IEEE
Transactions on Information Theory, \textbf{47}, no. 7, 3065 (2001).                                         

\bibitem{Hamada} M. Hamada, Proceedings IEEE International Symposium on
Information Theory, 480 (2003).   

\bibitem{Rains} E. M. Rains, IEEE Transactions on Information
Theory, \textbf{45}, no. 6, 1827 (1999).                                                                                                                    
\bibitem{Grassl} M. Grassl,  T. Beth, and M. Roetteler, International Journal of Quantum Information, \textbf{2}, no. 1, 55, (2004).

\bibitem{CWS1} A. Cross, G. Smith, J. Smolin, and B. Zeng,  arXiv: 0708.1201.

\bibitem{CWS2} I. Chuang, A. Cross, G. Smith, J. Smolin, and B. Zeng, arXiv: 0803.3232.

\bibitem{Markus1} M. Grassl and M. Roetteler, arXiv:0801.2150.

\bibitem{Markus2} M. Grassl and M. Roetteler, arXiv:0801.2144.

\bibitem{Yu2} S. Yu, Q. Chen, and C. H. Oh,  arXiv: 0709.1780.

\bibitem{Arvind} V. Arvind, Piyush P Kurur, K.R. Parthasarathy,
arXiv: quant-ph/0210097 (2002).

\bibitem{Calderbank} V. Aggarwal and R. Calderbank,
IEEE Trans. Inf. Theory, \textbf{54}, no. 4, 1700-1707, (2008).

\bibitem{Yu3} D. Hu, W. Tan, M. Zhao, Q. Chen, S. Yu, and C. H. Oh,  arXiv: 0801.0831.

\bibitem{Looi} S. Y. Looi, L. Yu, V. Gheorghiu, and R. B. Griffiths,  arXiv: 0712.1979.

\bibitem{Bar1} S. D. Bartlett, H. de Guise, and B. C. Sanders, Phys. Rev. A
\textbf{65}, 052316 (2002).

\bibitem{San} B. C. Sanders, S. D. Barlett, and H. de Guise,
Proceedings, ICSSUR'01, ed. D. Han, Y. S. Kim, B. E. A. Saleh, A. V. Sergienko, and M. C. Teich (2002).

\bibitem{Dab} J. Jamil, X. G. Wang, and B. C. Sanders, J. Phys. A: Math. Gen. \textbf{36} (14), 2525-2536 (2003).

\bibitem{Got} D. Gottemann, A. Kitaev and J. Preskill, Phys. Rev. A \textbf{65
}, 044303 (2002).

\bibitem{Pat} J. Patera and H. Zassenhaus, J. Math. Phys. \textbf{29}, 665
(1988).

\bibitem{Kni} E. Knill, arXiv: quant-ph/9608048 (1996).

\bibitem{Moor3} E. Hostens, J. Dehaene, and B. De Moor, Phys. Rev. A \textbf{71}, 042315 (2005).

\bibitem{Nielsen} M. A. Nielsen and I. L. Chuang, \textit{Quantum Computation and Quantum Information}, Cambridge University Press, Cambridge, UK, (2000).

\bibitem{S2} D. Schlingemann, arXiv: quant-ph/0111080 (2001).

\bibitem{knill-laflamme-theory} E.~Knill and R.~Laflamme, Phys.\ Rev.\ A {\bf 55}, 900
(1997).

\bibitem{bennett-tome} C.~Bennett, D.~DiVincenzo, J.~Smolin, and W.~Wootters, Phys.\
Rev.\ A {\bf 54}, 3824 (1996).

\end{thebibliography}
\end{document}